\documentclass[twocolumn,nofootinbib]{revtex4}

\usepackage{graphicx}

\begin{document}

\title{Introduction to Spin-Polarized Ballistic Hot Electron Injection and Detection in Silicon}

\author{Ian Appelbaum}
\affiliation{Center for Nanophysics and Advanced Materials and Department of Physics, University of Maryland, College Park MD 20742 USA}

\begin{abstract}
Ballistic hot electron transport overcomes the well-known problems of conductivity and spin lifetime mismatch that plagues spin injection in semiconductors with ferromagnetic ohmic contacts. Through the spin-dependent mean-free-path, it also provides a means for spin detection after transport. Experimental results using these techniques (consisting of spin precession and spin-valve measurements) with Silicon-based devices reveals the exceptionally long spin lifetime and high spin coherence induced by drift-dominated transport in the semiconductor. An appropriate quantitative model that accurately simulates the device characteristics for both undoped and doped spin transport channels is described; it can be used to determine the spin current velocity, diffusion constant, and spin lifetime, constituting a spin ``Haynes-Shockley'' experiment without time-of-flight techniques. A perspective on the future of these methods is offered as summary.  

\vspace{12pt}

\noindent KEYWORDS: spin polarized electrons, ballistic hot electron transport, spin injection and detection, semiconductor spintronics, spin precession
\end{abstract}

\maketitle

\section{\label{INTROSEC}Background}

The problems of spin-polarized electron injection and detection are central to the field of semiconductor spintronics. Since ferromagnetic metals have a large asymmetry between the spin-up and spin-down density of states and therefore high spin-polarization at the Fermi level, it is natural to assume that ohmic transport of these electrons into a semiconductor will provide a robust injection mechanism for spin-polarized currents. However, in the diffusive regime required by ohmic transport (where the electron mean free path (mfp) is smaller than all other lengthscales and an electron temperature and chemical potential are well-defined), injection and detection efficiencies are suppressed by the large differences in conductivity and spin lifetimes between metals and semiconductors, resulting in the so-called ``fundamental obstacle to spin injection''\cite{JOHNSON1988, SCHMIDT1, SCHMIDT2, RASHBA1}. 

The problem can be simply viewed in the following way: Since the spin-up and spin-down conductivities in the (non-magnetic) semiconductor are identical, a spin-polarized current must be driven by an asymmetric spin-dependent electrochemical potential ($\mu_{\uparrow / \downarrow}$) drop across the semiconductor. This necessarily requires $\Delta \mu = \mu_\uparrow - \mu_\downarrow \neq 0$ at the interface between the ferromagnet (FM) and semiconductor. In terms of the semiconductor bulk conductivity $\sigma_{S}$, the semiconductor spin transport lengthscale $L$, cross-sectional area $A$, spin-polarization $P$, and total current $I$, Ohm's law gives 

\begin{equation}
\Delta \mu = 2P \cdot IR=2PI\frac{L}{\sigma_S A}.
\label{scside}
\end{equation}
  
\begin{figure}
\includegraphics[width=6cm,height=4cm]{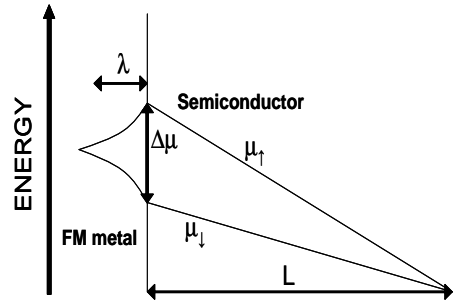}
\caption{\label{MUFIG} Spin-dependent band diagram showing the electrochemical potentials near the interface of a metallic FM (high conductivity and short spin lifetime) with a semiconductor (low conductivity and long spin lifetime).}
\end{figure}

On the other hand, in the metallic FM the spin current flowing toward the FM/semiconductor interface is caused instead by a spin-dependent conductivity $\sigma_{\uparrow / \downarrow}$. Therefore, the interfacial electrochemical splitting can also be written as 

\begin{equation}
\Delta \mu = I(R_\downarrow-R_\uparrow)=I \left( \frac{\lambda}{\sigma_\downarrow A}- \frac{\lambda}{\sigma_\uparrow A}  \right),
\label{fmside}
\end{equation}

\noindent where $\lambda$ is the lengthscale over which the non-equilibrium electrochemical splitting in the FM decays away from the interface. Since diffusive transport dominates in the metal because it cannot support internal electric fields, this lengthscale is the ``spin diffusion length'' $\sqrt{D \tau}$, where $\tau$ is the spin lifetime in the FM metal and $D$ is the diffusion coefficient. Unfortunately, this timescale is typically very short, so the lengthscale $\lambda$ is also short. The spin-dependent electrochemical potentials near the FM-semiconductor interface are schematically illustrated in Fig. \ref{MUFIG}. 

With the definition of the FM bulk polarization as $\beta=\frac{\sigma_\uparrow-\sigma_\downarrow}{\sigma_\uparrow+\sigma_\downarrow}$, we can rewrite the electrochemical splitting on the FM side (Eqn. \ref{fmside}) as

\begin{equation}
\Delta \mu = \frac{I\lambda}{\sigma_{FM} A}\frac{4\beta}{1-\beta^2},
\label{fmside2}
\end{equation}

\noindent where $\sigma_{FM}$ is the average value of conductivity $\frac{\sigma_\uparrow + \sigma_\downarrow}{2}$. At an ideal interface where the absence of interfacial spin relaxation preserves continuity of the electrochemical potentials,\cite{VANSON, JOHNSONREPLY} the expressions given for the interfacial electrochemical splitting in Eqns. \ref{scside} and \ref{fmside2} must be equal. This then allows us to calculate the polarization of the injected current in the semiconductor:

\begin{equation}
P=\frac{\sigma_S}{\sigma_{FM}} \frac{\lambda}{L} \left[ \frac{2\beta}{1-\beta^2}  \right].
\label{scpol}
\end{equation}
 
Note that the conductivity of FM metals is several orders of magnitude larger than any semiconductor, even in the degenerately doped regime. In addition, the spin diffusion length in the FM metal is many orders of magnitude shorter than the spin transport length in the semiconductor due primarily to the short (long) lifetime in the metal (semiconductor). The polarization of the electron current flowing through the semiconductor is then dependent on the product of two (very) small quantities, $\frac{\sigma_S}{\sigma_{FM}}$ and $\frac{\lambda}{L}$. Unless the bulk conductivity of the FM is close to $1$ (i.e. a 100\% polarized ferromagnetic half-metal -- still not possible in practice), ohmic injection of spin polarized electrons is therefore hopelessly inefficient. 

To circumvent this problem, the constraints of ohmic injection must be lifted. For instance, the requirement that the electrochemical splittings be equal on the FM and semiconductor side can be relaxed by inserting a resistive tunnel barrier.\cite{RASHBA1} The high conductivity of the FM can be addressed by using a ferromagnetic semiconductor which typically has a much lower carrier concentration than a metal.\cite{FIEDERLING, OHNO} 

Here, we consider a mechanism for spin injection that differs fundamentally from ohmic injection in that 

\begin{enumerate}
 \item{Momentum scattering is not the smallest lengthscale in the device; and}
 \item{Electrochemical potentials can not be uniquely defined in thermal equilibrium.}
\end{enumerate}

\noindent The above-described obstacle to spin injection therefore does not apply. With our techniques, the physical lengthscale of the metallic electron injection contacts is shorter than the mfp, and conduction occurs through states far above the Fermi level (as compared to the thermal energy $k_BT$), far out of thermal equilibrium. Because this transport mode utilizes electrons with high kinetic energy that do not suffer inelastic scattering, it is known as ``Ballistic Hot Electron Transport''. We will find that this transport mode is useful not only for spin-polarized electron injection into semiconductors, but also (and perhaps more importantly) for spin detection after transport through them.

\section{Hot Electron Generation and Collection}

\begin{figure}
\includegraphics[width=6cm,height=9cm]{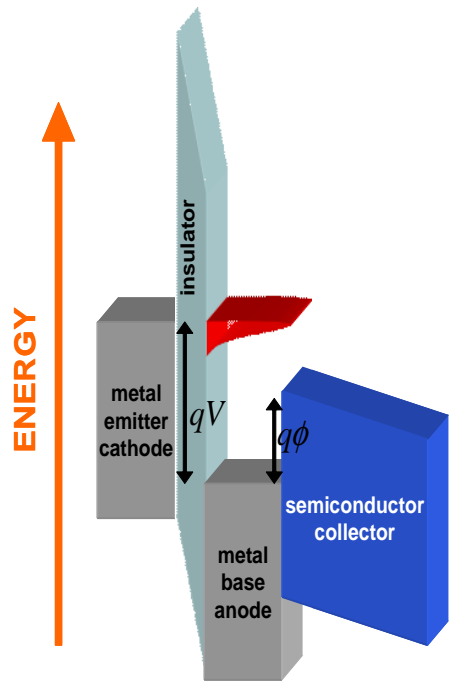}
\caption{\label{TJFIG} Schematic illustration of a tunnel junction used as a source for ballistic hot electron injection into a semiconductor conduction band. The emitter cathode electrostatic potential energy ($qV$) must exceed the Schottky barrier height $q\phi$.}
\end{figure}

Tunnel junctions (TJs, which consist of two metallic conductors separated by a thin insulator)\cite{BETHE, DUKE, FISHER, GIAEVER, SIMMONS, ROWELL} are the ideal hot electron source. The electrostatic potential energy $qV$ provided by voltage bias $V$ tunes the energy of hot electrons emitted from the cathode, and the exponential energy dependence of quantum-mechanical tunneling assures a narrow distribution. Hot electrons thus created can be collected by the Schottky barrier if $qV>q\phi$. Because of its robust insulating native oxide, Aluminum (Al) is often employed during tunnel junction fabrication; although in principle any insulator can be used, it has been empirically found that the best are Al$_x$O, MgO, and AlN.\cite{MOODERAMTJREVIEW, JMRBARRIERS} The first proposal for a tunnel-junction hot electron injector was by Mead, who suggested another tunnel junction or the vacuum level as a collector.\cite{MEAD1, MEAD2, MEADVACUUM} Very soon thereafter, Spratt, Schwartz, and Kane showed that a semiconductor collector could be used to realize this device with a Au/Al$_x$O/Al tunnel junction.\cite{SPRATT} A schematic band diagram of this device is shown in Fig. \ref{TJFIG}.

Metals have a very high density of electrons at and below the Fermi energy $E_F$. Therefore, sensitive hot electron collection by the semiconductor conduction band relies on an electrically rectifying barrier to eliminate transport of these thermalized electrons across the metal-semiconductor interface, which would dilute the injected hot electron current. This barrier is ideally created by the difference in work functions of the metal and the electron affinity of the semiconductor\cite{SCHOTTKY1, SCHOTTKY2, MOTT1, MOTT2}, but in reality its height is determined more by the details of surface states which lie deep in the bandgap of the semiconductor that pin the Fermi level. \cite{TERSOFF} There is of course always a leakage current due to thermionic emission over this ``Schottky'' barrier at nonzero temperature; because typical barrier heights are in the range 0.6-0.8 eV\cite{SZEBOOK} for the common semiconductors Si and GaAs, hot electron collection with Schottky barriers is often performed at temperatures below ambient conditions to reduce current leakage to negligible levels.

\section{Spin-Polarized Hot Electron Transport}

The mfp of hot electrons in ferromagnetic metal films is spin dependent: majority (``spin up'') electrons have a longer mfp than minority (``spin down'') electrons. Therefore, an initially unpolarized hot electron current will become spin-polarized by spin-selective scattering during ballistic transport through a ferromagnetic metal film, where the polarization is given by

\begin{equation}
P=\frac{e^{-\frac{l}{\lambda_{maj}}}-e^{-\frac{l}{\lambda_{min}}}}{e^{-\frac{l}{\lambda_{maj}}}+e^{-\frac{l}{\lambda_{min}}}},
\label{BSFEQN}
\end{equation}

\noindent where $l$ is the FM film thickness, $\lambda_{maj}$ is the majority spin mfp, and $\lambda_{min}$ is the minority spin mfp. This ``ballistic spin-filtering'' effect can be used not only for spin polarization at injection, but also for spin analysis of a hot electron current at detection, much as an optical polarizing filter can be used both for electric field polarization and analysis of photons by changing the relative orientation of the optical axis. The realization of this important spin analysis concept has enabled the spin injection, transport, and detection in Si described in this text.

\section{Spins in Silicon}

Silicon is a relatively simple spin system\cite{ZUTICNATURE}; Si could be considered as the ``hydrogen atom'' of semiconductor spintronics:

\begin{enumerate}
\item{A low atomic number ($Z=14$) leads to a reduced spin-orbit interaction (which scales as $Z^4$ in atomic systems).}
\item{As discussed previously, the preservation of spatial inversion symmetry of the diamond lattice leads to a spin-degenerate conduction band, eliminating D'yakonov-Perel' spin scattering in bulk Si.}
\item{The most abundant isotope of Si ($\approx 92$\%) is Si$^{28}$, which has no nuclear spin, and this abundance can be fortified with isotopic purification to very high levels.\cite{CARDONA} Therefore, the spin-nuclear (hyperfine) coupling is weak, as compared with other semiconductors where no such nuclear-spin-free isotope exists.}
\end{enumerate}

Due to its apparent advantages over other semiconductors, many groups tried to demonstrate phenomena attributed to spin transport in Si\cite{JIA, LEE, HWANG, HACIA1, HACIA2, DENNIS1, DENNIS2, DENNIS3, GREGG} but this was typically done with ohmic FM-Si contacts and two-terminal magnetoresistance measurements or in transistor-type devices\cite{TANAKA, CAHAY} which are bound to fail due to the ``fundamental obstacle'' for ohmic spin injection mentioned in Sec. \ref{INTROSEC}.\cite{SCHMIDT1, SCHMIDT2, RASHBA1, JOHNSON1988} Although weak spin-valve effects are often presented, no evidence of spin precession is available so the signals measured are ambiguous at best.\cite{MONZON, JOHNSON}  Indeed, although magnetic exchange coupling across ultra-thin tunneling layers of Si was seen, not even any spin-valve magnetoresistance was observed.\cite{GAREEV} 

These failures were addressed by pointing out that only in a narrow window of FM-Si Resistance-Area (RA) product was a large spin polarization and hence large magnetoresistance expected.\cite{FERTJAFFRES} Subsequently, several efforts to tune the FM-Si interface resistance were made.\cite{MIN1, MIN2, WANGCONTACT, UHRMANN, DIMOPOULOS} However, despite the ability to tune the RA product by over 8 orders of magnitude and even into the anticipated high-MR window (for instance, by using a low-work-function Gd layer), no evidence that this approach has been fruitful for Si can be found, and the theory has yet been confirmed only for the case of low-temperature-grown 5nm-thick GaAs.\cite{FERTMATTANA1, FERTMATTANA2}

Subsequent to the first demonstration of spin transport in Si,\cite{APPELBAUMNATURE} optical detection (circular electroluminescence analysis) was shown to indicate spin injection from a FM and transport through several tens of nm of Si, using first a Al$_x$O tunnel barrier\cite{JONKERNATPHYS} and then tunneling through the Schottky barrier\cite{KIOSEOGLOU}, despite the indirect bandgap and relative lack of spin-orbit interaction which couples the photon angular momentum to the electron spin. These methods required a large perpendicular magnetic field to overcome the large in-plane shape anisotropy of the FM contact, but later a perpendicular anisotropic magnetic multilayer was shown to allow spin injection into Si at zero external magnetic field.\cite{GRENET} While control samples with non-magnetic injectors do show negligible spin polarization, again no evidence of spin precession was presented. 

More recently, four-terminal nonlocal measurements on Si devices have been made, for instance with Al$_x$O\cite{JONKERLATERAL} or MgO\cite{SHIRAISHI} tunnel barriers, or Schottky contacts using ferromagnetic silicide injector and detector.\cite{ANDO} Only Ref. \onlinecite{JONKERLATERAL} presents evidence of spin precession, but the low signal-to-noise obscures geometrical effects of precession expected from the oblique magnetic field configuration used.\cite{OBLIQUE}

As mentioned in Sec. \ref{INTROSEC}, another way to overcome the conductivity/lifetime mismatch for spin injection is to use a carrier-mediated ferromagnetic semiconductor heterointerface. The interfacial quality may have a strong effect on the injection efficiency so epitaxial growth will be necessary. Materials such as dilute magnetic semiconductors Mn-doped Si\cite{LABELLA}, Mn-doped chalcopyrites\cite{IGORSIPRL, IGORNATMAT, KETTERSON1, KETTERSON2, CHALCOPYRITE}, or the ``pure'' ferromagnetic semiconductor EuO \cite{SCHLOM} have all been suggested, but none as yet have been demonstrated as spin injectors for Si. It should be noted that while their intrinsic compounds are indeed semiconductors, due to the carrier-mediated nature of the ferromagnetism, it is seen only in highly (i.e. degenerately) doped and essentially metallic samples in all instances.

\section{\label{DEVICESEC}Ballistic Hot Electron Injection and Detection Devices}

Two types of tunnel junctions have been employed to inject spin-polarized ballistic hot electrons into Si. The first used ballistic spin filtering of initially unpolarized electrons from a nonmagnetic Al cathode by a ferromagnetic anode base layer in direct contact to undoped, 10-micron-thick single-crystal Si(100).\cite{APPELBAUMNATURE} Despite SVT measurements suggesting the possibility of 90\% spin polarization in the metal,\cite{SVTREVIEW, VANDIJKEN3400} only 1\% polarization was found after injection into and transport through the Si. It was discovered later that a nonmagnetic Cu interlayer spacer could be used to increase the polarization to approximately 37\%,\cite{SPINFETEXPT} likely due to Si's tendency to readily form spin-scattering ``magnetically-dead'' alloys (silicides) at interfaces with ferromagnetic metals.

\begin{figure}
\includegraphics[width=6cm,height=6cm]{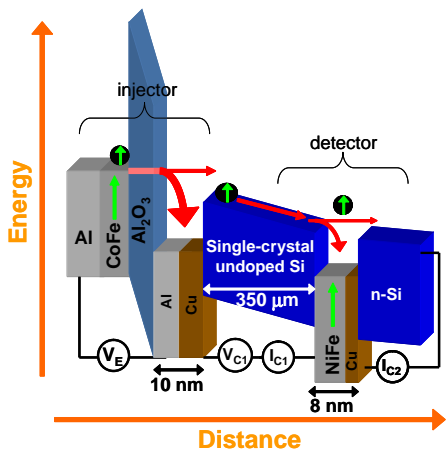}
\caption{\label{BANDDIAGRAMFIG} Schematic band diagram of a four-terminal (two for TJ injection and two for FM SMS detection) ballistic hot electron injection and detection device with a 350 $\mu$m-thick Si transport layer.}
\end{figure}

Despite the possibilities for high spin polarization with these ballistic spin filtering injector designs, the short hot-electron mfps in FM thin film anodes causes a very small injected charge current on the order of 100nA with an emitter electrostatic potential energy approximately 1eV above the Schottky barrier and a contact area approximately $100\times 100 \mu$m$^2$. Because the injected spin density and spin current are dependent on the product of spin polarization and charge current, this technique is not ideal for transport measurements. Therefore, an alternative injector utilizing a FM tunnel-junction cathode and nonmagnetic anode (which has a larger mfp) was used for approximately 10 times greater charge injection and hence larger spin signals, despite somewhat smaller potential spin polarization of approximately 15\%.\cite{35PERCENT, BIQINPRL} These injectors can be thought of as one-half of a magnetic tunnel junction,\cite{MOODERA} with a spin polarization proportional to the Fermi-level density of states spin asymmetry, rather than exponentially dependent on the spin-asymmetric mfp as is the case with ballistic spin filtering described above.

Although the injection is due to ballistic transport in the metallic contact, the mfp in the Si is only on the order of 10nm,\cite{BELLSIMFP} {\emph {so the vast majority of the subsequent transport to the detector over a lengthscale of tens,\cite{APPELBAUMNATURE, 35PERCENT, SPINFETEXPT} hundreds,\cite{BIQINPRL, OBLIQUE} or thousands\cite{2MM} of microns occurs at the conduction band edge following momentum relaxation}}. Typically, relatively large accelerating voltages are used so that the dominant transport mode is carrier drift; the presence of rectifying Schottky barriers on either side of the transport region assures that the resulting electric field does nothing other than determine the drift velocity of spin polarized electrons and hence the transit time\cite{CANALI, JACOBONI} -- there are no spurious (unpolarized) currents induced to flow. Furthermore, undoped Si transit layers are primarily used; otherwise band-bending would create a confining potential and increase the transit time, potentially leading to excessive depolarization (see Section \ref{DOPEDSEC}).\cite{DOPED} 

The ballistic hot electron spin detector is comprised of a semiconductor-FM-semiconductor structure (both Schottky interfaces), fabricated using UHV metal-film wafer bonding (a spontaneous cohesion of ultra-clean metal film surfaces which occurs at room-temperature and nominal force in ultra-high vacuum).\cite{SVT2}\cite{ALTFEDER} After transport through the Si, spin-polarized electrons are ejected from the conduction band over the Schottky barrier and into hot electron states far above the Fermi energy. Again, because the mfp in FMs is larger for majority-spin (i.e. parallel to magnetization) hot electrons, the number of electrons coupling with conduction band states in a n-Si collector on the other side (which has a smaller Schottky barrier height due to contact with Cu) is dependent on the final spin polarization and the angle between spin and detector magnetization. Quantitatively, we expect the contribution to our signal from each electron with spin angle $\theta$ to be proportional to:

\begin{eqnarray}
\cos^2{\frac{\theta}{2}}e^{-\frac{l}{\lambda_{maj}}}+\sin^2{\frac{\theta}{2}}e^{-\frac{l}{\lambda_{min}}}=\nonumber\\
\frac{1}{2}\left[ (e^{-\frac{l}{\lambda_{maj}}} - e^{-\frac{l}{\lambda_{min}}}) \cos{\theta}+(e^{-\frac{l}{\lambda_{maj}}} + e^{-\frac{l}{\lambda_{min}}})\right].
\end{eqnarray}

\noindent Because the exponential terms are constants, this has the simple form $\propto \cos{\theta}+const$; in the following, we disregard the constant term as it is spin-independent.

The spin transport signal is thus the (reverse) current flowing across the n-Si collector Schottky interface. In essence, this device (whose band diagram is schematically illustrated in Fig. \ref{BANDDIAGRAMFIG}) can be thought of as a split-base tunnel-emitter SVT with several hundred to thousands of microns of Si between the FM layers. 

\begin{figure}
\includegraphics[width=6cm,height=5cm]{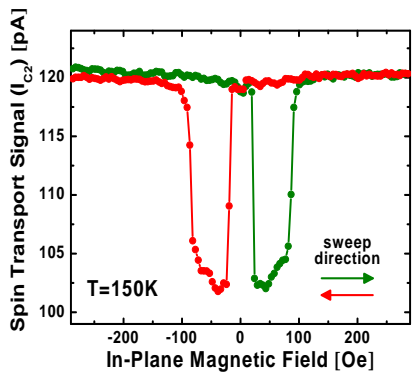}
\caption{\label{SPINVALVEFIG}In-plane magnetic field measurements show the ``spin-valve'' effect and can be used to calculate the spin polarization after transport.}
\end{figure}

\begin{figure}
\includegraphics[width=6cm,height=9cm]{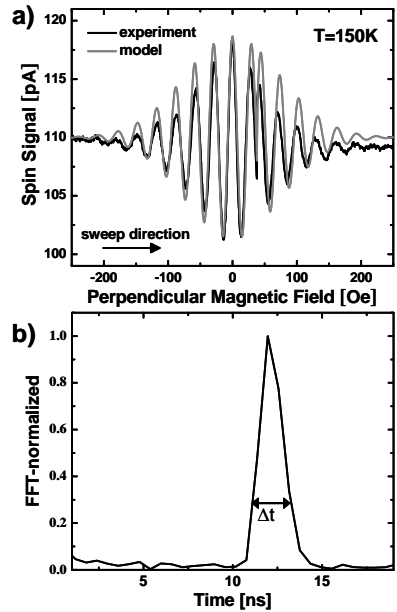}
\caption{\label{PRECESSIONFIG}(a) A typical spin precession measurement shows the coherent oscillations due to drift and the suppression of signal amplitude (``dephasing'') as the precession frequency rises. Our model simulates this behavior well. (b) The real part of the Fourier transform of the precession data in (a) reveals the spin current arrival distribution.}
\end{figure}

Two types of measurements are typically made: ``spin valve'' in a magnetic field parallel to the plane of magnetization, and spin precession in a magnetic field perpendicular to the plane of magnetization. The former allows the measurement of the difference in signals between parallel (P) and antiparallel (AP) injector/detector magnetization and hence is a straightforward way of determining the conduction electron spin polarization, 

\begin{equation}
P=\frac{I_P-I_{AP}}{I_P+I_{AP}}.
\label{PEQN}
\end{equation}

\noindent Typical spin-valve measurement data, indicating $\approx$8\% spin polarization after transport through 350$\mu$m undoped Si, is shown in Fig. \ref{SPINVALVEFIG}. 

Measurements in perpendicular magnetic fields reveal the average spin orientation after transit time $t$ through the Si, due to precession at frequency $\omega=g\mu_BB/\hbar$, where $B$ is magnetic field. If the transit time is determined only by drift, i.e. $t=\frac{L}{\mu E}$, we expect our spin transport signal to behave $\propto \cos{g\mu_BBt/\hbar}$. However, due to transit time uncertainty $\Delta t$ caused by random diffusion, there is likewise an uncertainty in spin precession angle $\Delta \theta=\omega \Delta t$ which increases as the magnetic field (and hence $\omega$) increases. When this uncertainty approaches $2\pi$ radians, the spin signal is fully suppressed by a cancellation of contributions from antiparallel spins, a phenomenon called spin ``dephasing'' or Hanle effect\cite{HANLE}.

On a historical note, our device is essentially a solid-state analogue of experiments performed in the 1950s that were used to determine the g-factor of the free electron in vacuum using Mott scattering as spin polarizer and analyzer and spin precession in a solenoid\cite{LOUISELL}. In our case, we already know the g-factor (from e.g. ESR lines), so our experiments in strong drift electric fields where spin dephasing is weak can be used to measure transit time with $t=h/g\mu_BB_{2\pi}$, where $B_{2\pi}$ is the magnetic field period of the observed precession oscillations, despite the fact that we make DC measurements, not time-of-flight\cite{CANALI, JACOBONI}.  Typical spin-precession data, indicating transit time of approximately 12ns to cross 350$\mu$m undoped Si in an electric field of $\approx$ 580 V/cm, is shown in Fig. \ref{PRECESSIONFIG}(a). 

\begin{figure}
\includegraphics[width=6cm,height=9cm]{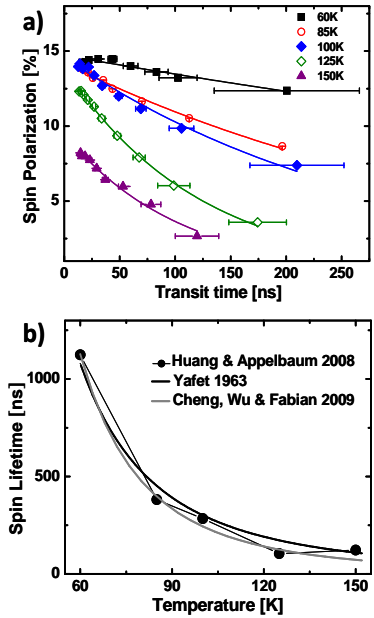}
\caption{\label{LIFETIMEFIG}(a) Fitting the normalized spin signal from in-plane spin-valve measurements to an exponential decay model using transit times derived from spin precession measurements at variable internal electric field yields measurement of spin lifetimes in undoped bulk Si. (b) The experimental spin lifetime values obtained as a function of temperature are compared to Yafet's T$^{-5/2}$ power law for indirect-bandgap semiconductors\cite{YAFET} and Cheng et al.'s T$^{-3}$ derived from a full bandstructure theory.\cite{FABIANWU}}
\end{figure}

One important application of this transit time information from spin precession is to correlate it to the spin polarization determined from spin-valve measurements and Eqn. \ref{PEQN} to extract spin lifetime. By varying the internal electric field, we change the drift velocity and hence average transit time. A reduction of polarization is seen with an increase in average transit time (as in Fig. \ref{LIFETIMEFIG}(a)) that we can fit well to first order using an exponential-decay model $P\propto e^{-t/\tau}$, and extract the timescale, $\tau$.\cite{BIQINPRL} In this way, we have observed spin lifetimes of approximately 1$\mu$s at 60K in 350$\mu$m-thick transport devices.\cite{BIQINTHESIS}\footnote{In Ref. \onlinecite{BIQINPRL}, a more conservative estimate of the spin lifetime (e.g. 520ns at 60K) was obtained by fitting to the transit time dependence of an alternative quantity expected to be proportional to the spin polarization, rather than using Eqn. \ref{PEQN} directly.}. The temperature dependence of spin lifetime is compared to the $T^{-5/2}$ power law predicted by Yafet\cite{YAFET} and the more recent (and more complete) theory of Cheng et al giving T$^{-3}$\cite{FABIANWU} in Fig. \ref{LIFETIMEFIG}(b). 

If we can adequately model the transport and the expected signals, we can make a spin-polarized electron version of the Haynes-Shockley experiment (originally used to measure diffusion coefficient, mobility, and lifetime of minority charge carriers).\cite{HAYNES} This experiment was very useful in the design of bipolar electronics devices such as junction transistors. By measuring the {\emph{spin}} transport analogues of these parameters, we imagine enabling the design of useful semiconductor spintronic devices.

\section{Spin Transport Modeling}

Modeling the spin transport signal involves summing all the detector signal contributions of spins which begin at the injector at the same time and with the same spin orientation (i.e. initial conditions on spin density $s(x,t=0)=\delta (x)$), where $s(x,t)$ is determined by the drift-diffusion equation 

\begin{equation} 
\frac{\partial s}{\partial t}=D\frac{\partial^2 s}{\partial x^2}-v\frac{\partial s}{\partial x}-\frac{s}{\tau},
\label{DRIFTDIFFEQN}
\end{equation} 

\noindent where $D$ is the diffusion coefficient and $v$ is the drift velocity. The solution to this partial differential equation with the $\delta(x)$ Dirac-delta initial condition is called the ``Green's function'', which can be used to construct the response to arbitrary spin injection conditions, including the DC currents used so far in experiments.

Unlike open-circuit voltage spin detection\cite{JOHNSON, LOU, JEDEMA}, our ballistic hot-electron mechanism employs current sensing. This must be accounted for in the modeling, and since electrons crossing the metal-semiconductor boundary do not return via diffusion, it imposes an absorbing boundary condition on the spin transport.\cite{APPELBAUMSTANESCU} Whereas voltage detection is sensitive to the spin density $s(x=L,t)$ and the boundaries imposed on the drift-diffusion Green's function are at infinity, here our signal is sensitive to the spin current $J_s(x=L,t)=-D\frac{ds}{dx}|_{x=L}$, where we must impose $s(x=L,t)=0$. 

\subsection{Undoped Si}

Assuming $v$ is a constant independent of position $x$ (as it is in undoped semiconductors), the Green's function of Eqn. \ref{DRIFTDIFFEQN} satisfying this boundary condition can be found straightforwardly using the method of images: 

\begin{equation}
s(x,t)=\frac{1}{2\sqrt{\pi Dt}} \left[ e^{-\frac{(x-vt)^2}{4Dt}}- e^{Lv/D} e^{-\frac{(x-2L-vt)^2}{4Dt}} \right]e^{-t/\tau}.
\label{GREENSEQ}
\end{equation} 

\noindent The corresponding spin current at the detector is 

\begin{equation}
J_s(x=L,t)=\frac{1}{2\sqrt{\pi Dt}}\frac{L}{t}e^{-\frac{(L-vt)^2}{4Dt}}e^{-t/\tau}.
\label{SPINCURREQ}
\end{equation}

\begin{figure}
\includegraphics[width=6cm,height=6cm]{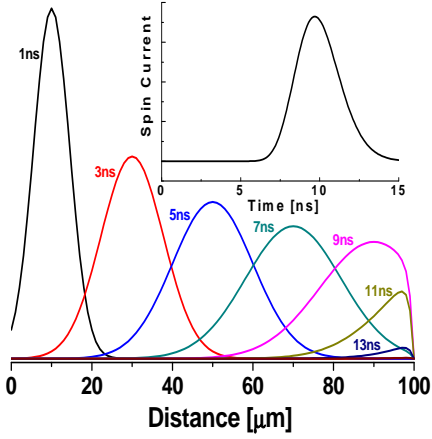}
\caption{\label{GREENSFIG} Evolution of a Delta function spin distribution (Green's function) injected on the left side at time increments shown in the legend, with drift velocity $10^6$cm/s, diffusion coefficient 100cm$^2$/s, and absorbing boundary conditions at the detector ($x=100\mu$m). Inset: Simulated spin current, given by the gradient of the Green's function at the detector, as in Eqn. \ref{SPINCURREQ}.
}
\end{figure}

\noindent The Green's function in Eqn. \ref{GREENSEQ} and the spin current derived from it in Eqn. \ref{SPINCURREQ} (using typical values for $v$, $D$, and $L$ at various times $t$) are shown in Fig. \ref{GREENSFIG}.

Note that the expression in Eqn. \ref{SPINCURREQ} is simply the spin density in the absence of the boundary condition\cite{JEDEMA, LOU} multiplied by the spin velocity $L/t$. Because our measurements are under conditions of strong drift fields, we have approximated this term with a constant (e.g. the average drift velocity $v$) in previous work.\cite{SPINFETTHRY, DEPHASING, BIQINPRL} To first order, this approximation involves a rigid shift of the signal peak to lower arrival times of $\approx 2D/v^2$ and a subsequent error in the measured velocity of $2D/L$. There is no first-order error in the dephasing. Because typical values for these variables are on the order of $v >10^6$ cm/s, $L>10^{-2}$cm, and $10^2<D<10^3$cm$^2$/s, the relative error associated with this approximation ($2D/Lv$ in both cases) is then just a few percent.

We can use this Green's function in the kernel of an integral expression to model, for instance, the expected precession signal as the weighted sum of all the cosine contributions from electrons with different arrival times ($0<t<\infty$):

\begin{equation} 
\int_{0}^\infty J_s(x=L,t) \cos{\omega t} dt.
\label{PRECESSIONINTEGRALEQN}
\end{equation} 

\noindent Note that because $J_s$ is causal (i.e. $J_s(t<0)=0$), we can extend the lower integration bound to $-\infty$ so that this expression is equivalent to the real part of the Fourier transform of $J_s(x=L,t)$. We can therefore use precession measurements as a direct probe of the spin current transit time distribution by making use of the Fourier transform, as shown in Fig. \ref{PRECESSIONFIG}(b). By fitting the data to the simulation, we can extract the transport variables $v$ and $D$; along with the spin lifetime measurement, this comprises a spin Haynes-Shockley experiment.

Eqn. \ref{PRECESSIONINTEGRALEQN} is valid only for the case of a purely perpendicular magnetic field $\vec{B}=B_z\hat{z}$.\cite{MOTSNYI} In the most general case, if there is also an in-plane field component $B_y\hat{y}$ (due to misalignment or a static field) and/or if the perpendicular component of the field $B_z$ is strong enough to overcome the in-plane magnetic anisotropy of the film, the $\cos{\omega t}$ term must be replaced with

\begin{eqnarray}
\frac{(B_z\cos\theta_1+B_y\sin\theta_1)(B_z\cos\theta_2+B_y\sin\theta_2)}{B_y^2+B_z^2}\cos{\omega t}\nonumber \\
+\frac{(B_z\sin\theta_1-B_y\cos\theta_1)(B_z\sin\theta_2-B_y\cos\theta_2)}{B_y^2+B_z^2},
\label{OBLIQUEEQN}
\end{eqnarray}

\noindent where in the simplest case the magnetization rotation angles $\theta_{1,2}=\arcsin(\tanh{(B_z/\eta_{1,2})})$, and $\eta_{1,2}$ are the demagnetization fields of injector and detector ferromagnetic layers (typically several Tesla).\cite{DOPED} From Eqn. \ref{OBLIQUEEQN}, we can show that small misalignments from the out-of-plane direction during single-axis precession measurements cause only in-plane magnetization switching of injector and detector and a negligible correction to the amplitude of precession oscillations to first order.\cite{OBLIQUE} Furthermore, the expression is now composed of a coherent term proportional to $\cos{\omega t}$ and an incoherent term independent of $t$. In the case of extremely strong dephasing (caused e.g. by geometrically-induced transit length uncertainty), the first term contributes a negligible amount to the integral expression in Eqn. \ref{PRECESSIONINTEGRALEQN} and the signal is dominated by the field and magnetization geometry dictated in the second term.\cite{2MM}

\subsection{\label{DOPEDSEC}Doped Si}

\begin{figure}
\includegraphics[width=6cm,height=6cm]{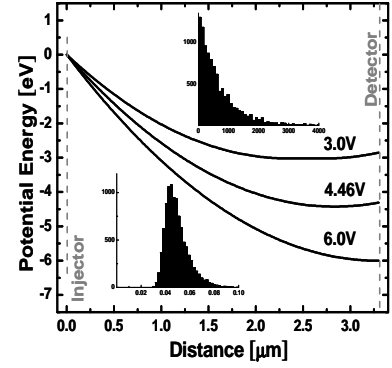}
\caption{\label{DOPEDFIG}Depletion-approximation conduction-band diagrams of a 3.3$\mu$m-thick n-type ($7.2\times 10^{14}$cm$^{-3}$) doped Si spin-transport layer with injector-detector voltage drop of 3V (where the transport is dominated by diffusion against the electric field at the detector side and results in the exponential-like arrival distribution with typical timescale 1000ns shown in the top inset); 4.46V (where the bias is enough to eliminate the electrostatically neutral region and fully deplete the transport layer; and 6V (where the potential well has been eliminated and transport is dominated by drift in the unipolar electric field, resulting in the gaussian-like arrival distribution with typical timescale of 0.050ns shown in the bottom inset).}
\end{figure}

In the above, we have assumed that the drift velocity $v$ in Eqn. \ref{DRIFTDIFFEQN} was independent of $x$. This assumption is not valid in doped semiconductors due to the presence of inhomogeneous electric fields and band-bending caused by ionized dopants. Eqn. \ref{DRIFTDIFFEQN} then no longer has constant coefficients, so in general its Green's function cannot be solved analytically and a computational approach must be taken. 

Direct simulation of an ensemble of electrons can be used to assemble a histogram of transit times, which in the limit of large numbers of electrons yields the transit-time distribution function. This ``Monte Carlo'' method, which incorporates the proper boundary conditions for our current-based spin detection technique automatically, was used to model spin transport through doped Si.\cite{DOPED} 

The technique involves discretely stepping through time a duration $\delta t$, modeling drift with a spatial translation $v(x) \delta t$, and diffusion with a translation of $\pm\sqrt{2D\delta t}$ (where the sign is randomly chosen to model the stochastic nature of the process), until $x>L$. The drift velocity $v(x)$ can be constructed from a depletion-approximation model of the transport layer to determine the electric field profile $E(x)$ and a mobility model to get $v(E)$.\cite{JACOBONI} Our experimental studies with n-type ($\approx 10^{15} cm^{-3}$ Phosphorus) 3.3$\mu$m-thick Si devices indicate that despite highly ``non-ohmic'' spin transport where the transit time can be varied by several orders of magnitude resulting from a confining electrostatic potential, Monte-Carlo device modeling shows that spin lifetime is not measurably impacted from its value in intrinsic Si.\cite{DOPED} Fig. \ref{DOPEDFIG} shows example band diagrams under different bias conditions (calculated in the depletion approximation) and corresponding arrival time distributions from the Monte-Carlo simulation.

\section{Discussion}

There has been significant progress in using ballistic hot electron spin injection and detection techniques for spin transport studies in Si. However, there are limitations of these methods. For example, the small ballistic transport transfer ratio is typically no better than $10^{-3}-10^{-2}$; the low injection currents and detection signals obtained will result in sub-unity gain and limit direct applications of these devices. In addition, our reliance on the ability of Schottky barriers to serve as hot electron filters presently limits device operation temperatures to approximately 200K (although materials with higher Schottky barrier heights could extend this closer to room temperature), and also limits application to only non-degenerately doped semiconductors. Carrier freeze-out in the n-Si spin detection collector at approximately 20K introduces a fundamental low-temperature limit as well.\cite{DEPHASING} 

Despite these shortcomings, there are also unique capabilities afforded by these methods, such as independent control over internal electric field and injection current, and spectroscopic control over the injection energy level\cite{BIQINJAP}. Unlike, for instance, optical techniques, other semiconductor materials should be equally well suited to study with these methods. The purpose is to use these devices as tools to understand spin transport properties for the design of spintronic devices just as the Haynes-Shockley experiment enabled the design of electronic minority-carrier devices.

There is still much physics to be done with ballistic hot electron spin injection and detection techniques applied to Si. Recently, we have adapted our fabrication techniques to assemble lateral spin transport devices, where in particular very long transit lengths\cite{2MM} and the effects of an electrostatic gate to control the proximity to a Si/SiO$_2$ interface can be investigated.\cite{LATERAL} Hopefully, more experimental groups will develop the technology necessary to compete in this wide-open field. Theorists, too, are eagerly invited to address topics such as whether this injection technique fully circumvents the ``fundamental obstacle'' because of a remaining Sharvin-like effective resistance,\cite{RASHBA2} or whether it introduces anomalous spin dephasing.\cite{DEPHASING}

\begin{acknowledgments}
Biqin Huang, Hyuk-Jae Jang, Jing Li, and Jing Xu provided invaluable fabrication, measurement, and programming efforts which have made much of this work possible.

This work has been financially supported by the Office of Naval Research, DARPA/MTO, and the National Science Foundation.
\end{acknowledgments}

\end{document}